\documentclass[amsmath,twocolumn,amssymb,floatfix,showpacs,
superscriptaddress,nofootinbib,longbibliography]{revtex4-2}
\usepackage{mathtools}
\usepackage{braket}
\usepackage[dvipsnames]{xcolor}
\usepackage{float}
\usepackage{subfigure}
\usepackage{graphics}
\usepackage{bm}
\usepackage{MnSymbol}
\usepackage{braket}
\usepackage{float}
\usepackage{tikz}
\usepackage{blkarray}
\usepackage{pgffor}
\usepackage{float}
\usepackage{silence}
\usepackage[colorlinks=true,linktoc=page,linkcolor=BrickRed,citecolor=magenta,urlcolor=purple]{hyperref}

\mathchardef\mhyphen="2D 

\newcommand\bea{\begin{eqnarray}}
\newcommand\eea{\end{eqnarray}}
\newcommand\beq{\begin{equation}}  
\newcommand\eeq{\end{equation}}

\usepackage[normalem]{ulem}
\definecolor{lime}{HTML}{A6CE39}
\usepackage{sidecap,tikz}
\usepackage{enumitem}
\DeclareRobustCommand{\orcidicon}{\hspace{-1.0mm}
	\begin{tikzpicture}
		\draw[lime, fill=lime] (0.0,0.0) 
		circle [radius=0.15] 
		node[white] {{\fontfamily{qag}\selectfont \tiny \,ID}};
		\draw[white, fill=white] (-0.0525,0.095) 
		circle [radius=0.007];
	\end{tikzpicture}
	\hspace{-3.0mm}
}
\foreach \x in {A, ..., Z}{\expandafter\xdef\csname orcid\x\endcsname{\noexpand\href{https://orcid.org/\csname orcidauthor\x\endcsname}
		{\noexpand\orcidicon}}
}

\begin{document}
\title{
Controlling Dissipative Topology Through Floquet Driving: From Transient Diagnostics to Boundary States Isolation
}
\author{Koustav Roy$^\S$}
\email[]{koustav.roy@iitg.ac.in} 
\affiliation{Department of Physics, Indian Institute of Technology Guwahati-Guwahati, 781039 Assam, India}
\author{Shahroze Shahab
$^\S$
}
\email[]{shahroze3141@gmail.com} 
\affiliation{Department of Physics, National Institute of Technology, Sector 1, Rourkela, Odisha 769008, India}
\author{Saurabh Basu}
\affiliation{Department of Physics, Indian Institute of Technology Guwahati-Guwahati, 781039 Assam, India}


\def\thefootnote{$\S$}\footnotetext{These authors contributed equally to this work}

\begin{abstract}
Engineering dissipative dynamics in open quantum systems is under active focus, especially in topological settings where resilient edge modes are expected to exhibit decay rates distinct from the bulk. In this letter, we propose an efficient dynamical scheme to discern such long-lived excitations. Employing a Floquet–Lindblad framework, we explore how periodic driving reshapes the key features of a paradigmatic topological model, namely a Creutz ladder. Our results bear testimony to a drive-induced \textit{unipolar-bipolar} transition in the Liouvillian skin effect, which gets dynamically manifested as a \textit{chiral-helical} damping crossover. Such a transition effectively rescales the bulk localization length, giving rise to a \textit{polarization drift} that we identify as a new invariant for efficient diagnosis of the nontrivial phases. As the transition becomes more gradual via tuning drive-rescaled parameters, we uncover signatures of a scale-free localization where skin and extended modes coexist with distinct decay rates. The emergent hierarchy of the decay rates yields two disparate timescales: a chiral wavefront that rapidly empties the bulk, followed by a long-lived regime dominated by robust edge modes. Overall, our results provide convincing evidence that periodic driving serves as a powerful handle to manipulate dissipative topological phases and dynamically isolate the boundary modes.
\end{abstract}

\maketitle
\textit{Introduction.-}
Advances in the control of dissipation and quantum coherence have reignited interest in the physics of open and nonequilibrium quantum systems \cite{advances_open_systems1,advances_open_systems2}. In this context, effective non-Hermitian (NH) descriptions have emerged as powerful tools to capture the dynamics of particle non-conserving systems \cite{NH_theory1,NH_theory2,NH_theory3,NH_theory4,NH_theory5,NH_theory6}. The time evolution of such systems is governed by the Gorini–Kossakowski–Sudarshan–Lindblad (GKSL) master equation \cite{LME1,LME2,LME3}, which unifies coherent and dissipative processes within a single Liouvillian framework, providing a natural arena to explore NH topology in the context of open quantum dynamics.
\par Building on this connection, recent studies have demonstrated that hallmark NH features, such as the non-Hermitian skin effect (NHSE) \cite{NHSE4,NHSE5,NHSE6,NHSE7}, exceptional points \cite{EP1,EP2,EP3,EP4}, and symmetry-protected amplification \cite{NH10fold,NH10foldfloquet} etc., can manifest in Liouvillian spectra as well, leading to the so-called Liouvillian skin effect (LSE) \cite{dissipative_topology1,dissipative_topology2,dissipative_topology3,dissipative_topology4,dissipative_topology5,dissipative_topology6,dissipative_topology7}. These discoveries not only extend topological concepts to dissipative systems, but also open routes toward applications in photonic and acoustic platforms, including unidirectional lasing \cite{LME_application1}, topological light transport \cite{LME_application2, LME_application3}, and enhanced sensing near exceptional degeneracies \cite{LME_application4,LME_application5,LME_application6}.
\par Despite these advances, several fundamental issues remain unresolved. In particular, the dynamical role of boundary modes in approaching the nonequilibrium steady state, and their possible isolation from skin-localized bulk modes, remain largely unexplored. while Liouvillian dynamics can, in principle, support chiral transport, where information flow is unidirectional \cite{LME_application1,information_constraint1,information_constraint2}, a precise and active control over this propagation is still in a state of its infancy.
\begin{figure}[t]
         \includegraphics[width=0.95\columnwidth]{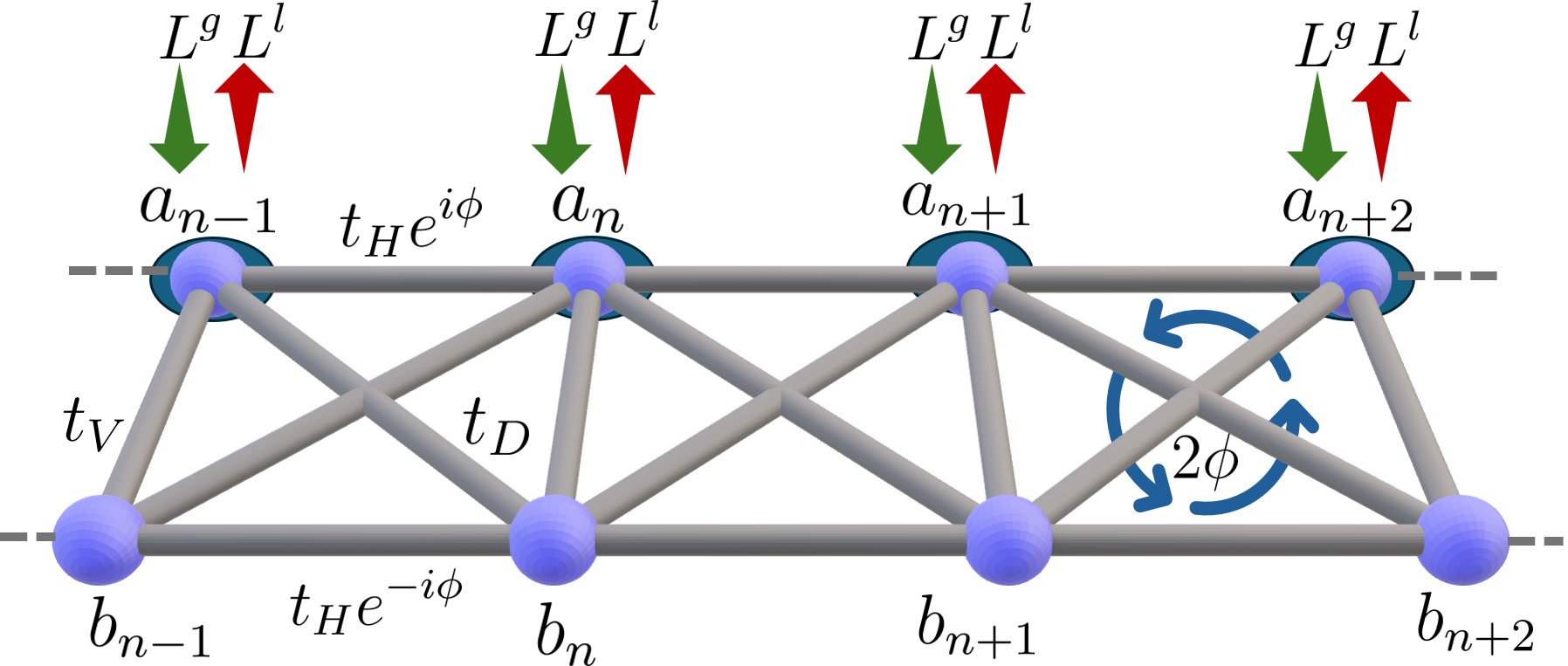}
\caption{Schematic representation of a Creutz ladder (weakly) coupled to an environment, with dissipation acting solely on the $A$ sublattice edge.
}
\label{fig:1}
\end{figure}
\par Meanwhile, periodic driving offers a versatile means to control NH effects far from equilibrium \cite{floquetformalism1,floquetformalism2,floquetformalism3,floquet1,floquet2,floquet3,floquet4}. The interplay between Floquet dynamics and NH topology has revealed unconventional phases and tunable decay dynamics \cite{GBZ_SSH_Floquet1,GBZ_SSH_Floquet2,gong_NH_ssh,NHCreutz2,Roy_Floquet_Creutz1,Roy_Floquet_Creutz2}, raising a broader question: can periodic driving dynamically control localization length and decay rates to isolate boundary states from skin-localized bulk modes during transient evolution?
Remarkably, periodic driving enables precisely such control, facilitating state-selective decay, directional information transfer, and alleviation of dissipative constraints. In this Letter, we address these aspects by developing an exact Floquet–Lindblad framework that captures the interplay between periodic driving, dissipation, and topology in open quantum systems.
\par Collectively, our results demonstrate a dynamical route for synthesizing robust topological edge modes via a drive-induced polarization drift that identifies the underlying topology; this is realized through controlled tuning of the driving parameters to ensure state-selective decay, thereby triggering a rapid chiral wavefront that empties the bulk and leaves behind the protected boundary modes as long-lived excitations.
\par \textit{Creutz Ladder in Floquet-Lindblad framework.-} Our starting point is a paradigmatic model for realizing chiral fermions, namely the quasi-1D Creutz ladder \cite{creutz1,creutz2,creutzexpt1,creutzexpt2}, composed of two coupled chains connected via horizontal ($t_H$), vertical ($t_V$), and diagonal ($t_D)$ tunneling amplitudes.
A magnetic flux $\phi$ threads each plaquette through a Peierls phase in the horizontal hopping. In momentum space, the Hamiltonian read as,
\begin{align} H(k)=2t_{H}[\cos{k}\cos{\phi}\,\sigma_0&+\sin{k}\sin{\phi}\,\sigma_z]\nonumber\\&+(t_V+2t_{D}\cos{k})\,\sigma_x. \label{eq:2} \end{align}
\begin{figure}[t]
         \includegraphics[width=\columnwidth]{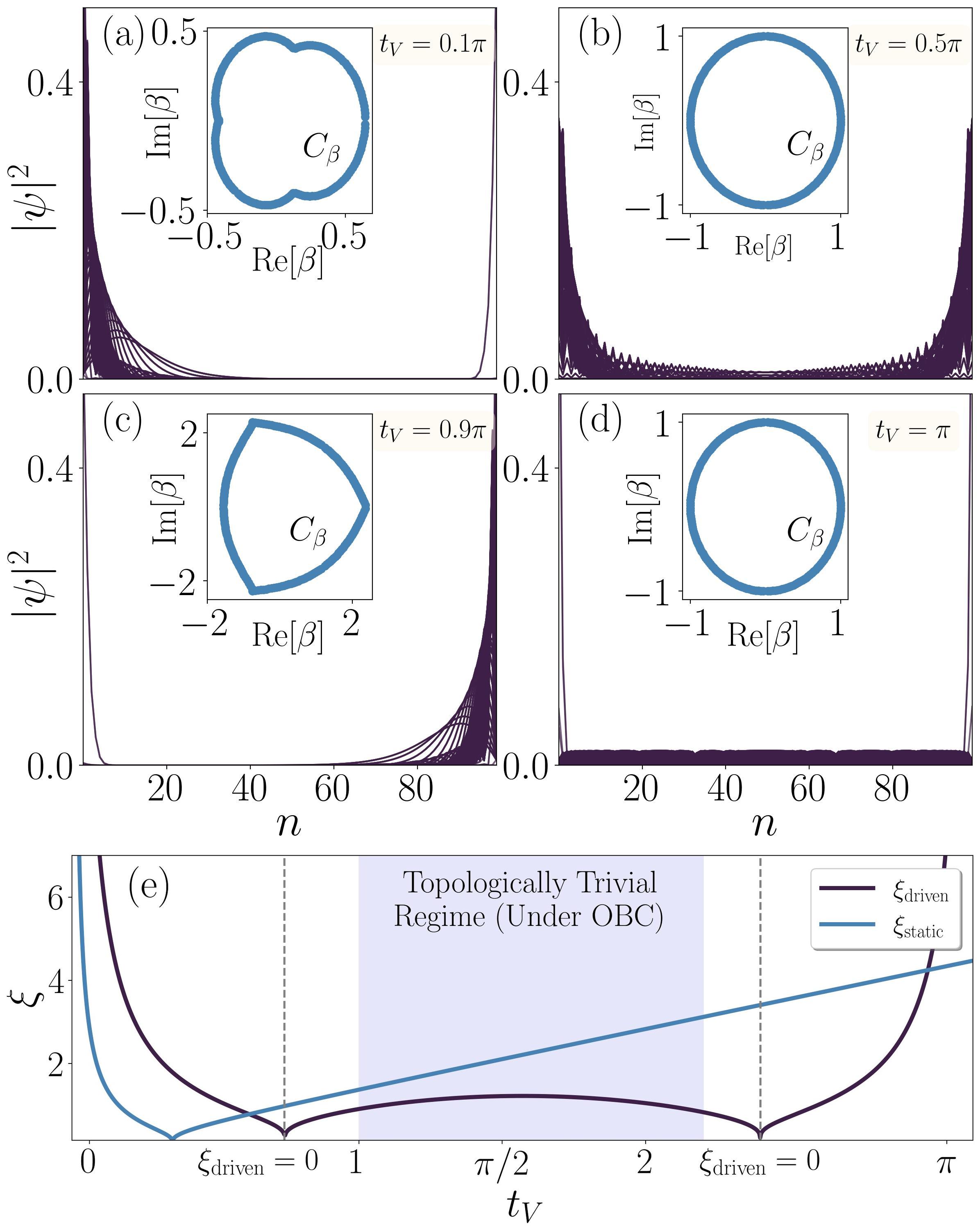}
\caption{(a–d) Representative eigenstate profiles for selected values of $t_V$. 
Unipolar localization with FGBZ radius $<1$ evolves into a bipolar configuration as $t_V$ increases, 
followed by skin suppression associated with a reversal of eigenstate polarization. 
(e) Drive-renormalized localization length, $\xi_{\text{driven}}$ showing periodic modulation and divergences at the skin-suppression points. Here we have chosen $L=100, t_D=t_H=0.5,\gamma_l=\gamma_g=0.3$, and $T=2$.
}
\label{fig:2}
\end{figure}
Moreover, for the $\pi$-flux configuration ($2\phi = \pi$), the ladder supports a complete set of symmetries essential for realizing nontrivial phases. As shown in Fig. \ref{fig:1}, the ladder is weakly coupled to a Markovian bath, leading to the following form of Lindblad master equation \cite{LME1,LME2,LME3},
\begin{equation}
\frac{d\rho}{dt} = -i[H, \rho] + \sum_{\alpha} \left( L_\alpha \rho L_\alpha^{\dagger} - \frac{1}{2} \left\{ L_\alpha^{\dagger} L_\alpha, \rho \right\} \right).
\end{equation}
where the dissipative processes are assumed to act only on the $A$-sublattice through a single loss and gain channel, 
\(
L_{x}^{l} = \sqrt{\gamma_{l}}\, c_{x,A}
\)
and 
\(
L_{x}^{g} = \sqrt{\gamma_{g}}\, c_{x,A}^{\dagger}.
\) At this point, the transient topology of the system, prior to significant quantum jump activity, is governed by an effective NH Hamiltonian, \( H_{\text{eff}} = H - i \sum_{\alpha} L_{\alpha}^{\dagger} L_{\alpha} \) which for $\phi=\pi/2$, becomes, \begin{equation}
    H_{\text{eff}}(k) = (t_V + 2t_D \cos k)\, \sigma_x + (2t_H \sin k + i\gamma/2)\, \sigma_z - i\gamma/2\, \mathbb{I},
\end{equation}
with $\gamma = \gamma_l + \gamma_g$. This form of the Hamiltonian is known to exhibit NHSE due to asymmetric gain–loss contributions \cite{NHCreutz2,Roy_Floquet_Creutz1}. Finally, to capture the evolution of the system, it is convenient to monitor the single-particle correlation matrix \cite{dissipative_topology1,dissipative_topology2,dissipative_topology3}, \( C_{ij} = \mathrm{Tr}\!\left[ c_{i}^{\dagger} c_{j} \rho(t) \right] \),
whose dynamics are analyzed in terms of its deviation from the stationary steady state, $\dot{C}_{SS}$, via \( \tilde{C}(t) = C(t) - C_{\text{SS}} \), which evolves according to, \( \tilde{C}(t) = e^{H_{\text{eff}} t}\, \tilde{C}(0)\, e^{H_{\text{eff}}^{\dagger} t} \). 
\par However, the methodology outlined above proves inadequate under periodic driving. Since the time-dependent Liouvillian prevents the steady-state solution from being obtained through the continuous Lyapunov equation, as was done in the static limit \cite{dissipative_topology1,dissipative_topology2,dissipative_topology3,dissipative_topology4,dissipative_topology5,dissipative_topology6,dissipative_topology7}, additionally, a NH analogue of $H_{\text{eff}}$ cannot be straightforwardly defined. To overcome these challenges, we employ the \textit{third quantization} framework \cite{third_quantization1,third_quantization2,third_quantization3} which proceeds as follows.
Starting with $n$ complex fermions we define 2$n$ Majorana operators constructed via $c_j = (w_{2j-1} - i w_{2j}) /2$, such that, 
\( H = \sum_{p,q} w_{p}^{\dagger} \mathcal{H}_{pq} w_{q}\), and  
\( L_{\alpha} = \sum_{p} l_{p}^{\alpha} w_{p} \). The Liouvillian can then be expressed in a quadratic form in the Fock space of superoperators, $\hat{\phi}_j^\dagger$, $\hat{\phi}_j$, as,
\begin{align}
\mathcal{L} 
= \frac{1}{2} 
\begin{pmatrix} 
\hat{\phi}^\dagger & \hat{\phi} 
\end{pmatrix}
\begin{pmatrix} 
- \mathbf{X}^\dagger & -i\mathbf{Y} \\ 
0 & \mathbf{X} 
\end{pmatrix}
\begin{pmatrix} 
\hat{\phi} \\ 
\hat{\phi}^\dagger 
\end{pmatrix}
- \frac{1}{2} \mathrm{Tr}(\mathbf{X}).
\label{eq:liouvillian_matrix}
\end{align}
The dynamics are primarily governed by the so-called damping matrix, \( \mathbf{X} = -4i\, \mathbf{H} + 2(\mathbf{M} +  \mathbf{M}^{\mathsf{T}})  \), with \( \mathbf{M}_{pq} = \sum_{\alpha} l_{p}^{\alpha} \, {l_{q}^{\alpha}}^{*} \). Importantly, the matrix $\mathbf{X}$ dictates the approach to the steady state, functioning as the dynamical counterpart to $H_{\text{eff}}$.
\par Within this representation, the Liouvillian acquires a compact matrix form that extends seamlessly to periodically driven open systems. Therefore, for a time-periodic Liouvillian $\mathcal{L}(t)$, the stroboscopic evolution over one period $T$ is given by, \( U_{\mathcal{L}} = \mathcal{T} \exp\!\left[ \int_{0}^{T} dt\, \mathcal{L}(t) \right] = e^{\mathcal{L}_F T} \), where $\mathcal{L}_F$ denotes the Floquet Liouvillian, expressed in the superoperator basis as,
\( e^{\mathcal{L}_F T} = 
\begin{pmatrix}
e^{-\mathbf{X_F}^{\dagger} T} & -i\mathbf{Q} \\
0 & e^{\mathbf{X_F} T}
\end{pmatrix} \), with $\mathbf{X_F}$ representing the Floquet damping matrix \cite{Floquet_Lindblad1,Floquet_Lindblad2,Floquet_Lindblad3}, that serves as the driven counterpart of $\mathbf{X}$ in the static case. To elucidate the implications of driving, we consider a step modulation protocol, where the damping matrix alternates between two configurations within one period, for example, 
\begin{equation}
\mathbf{X_k}(t) =
\begin{cases}
\mathbf{X_k}(d_x, d_z = 0), & \text{for } t \in [lT,\, lT + T/2), \\[0.5pt]
\mathbf{X_k}(d_x = 0, d_z), & \text{for } t \in [lT + T/2,\, lT + T).
\end{cases}
\end{equation}

\par \textit{Drive-induced chiral-to-helical damping.-}
Analogous to the static case, the Floquet Liouvillian, $\mathbf{X_F}$, retains the hallmark feature of the LSE (see Fig.~\ref{fig:2}(a) for $t_V = 0.1\pi$). The emergence of such nontrivial boundary modes is captured through the non-Bloch framework, where the conventional Brillouin zone is replaced by a generalized Brillouin zone (GBZ) parameterized by $\beta = e^{ik}$ \cite{GBZ1,GBZ2,GBZ3}. Unlike the static scenario, determination of the Floquet GBZ (FGBZ) requires symmetric time frames in which $X_F$ assumes a massless Dirac structure. Followed by which a second-order Taylor expansion of the characteristic equation yields a tractable form of the FGBZ (details in the Supplementary Material (SM)). Interestingly, analysis of the FGBZ reveals a series of special driving points, $t_V = 2n\pi/T$ and $t_V = (2n+1)\pi/T$ ($n \in \mathbb{Z}$), where the FGBZ takes the form of a unit circle. At $t_V = 2n\pi/T$, the skin effect is completely suppressed, while at $t_V = (2n+1)\pi/T$ the system exhibits \textit{bipolar localization}, with states equally distributed at both boundaries. For instance, tuning $t_V$ from $0.1\pi$ to $0.5\pi$ drives a transition from a unipolar to a bipolar configuration [Fig.~\ref{fig:2}(b)], where positive- (negative-) energy modes localize at the left (right) edge. This marks the emergence of an \textit{anomalous} $\mathbb{Z}_2$ \textit{skin effect} from an originally $\mathbb{Z}$-classified system \cite{NHSE5}. Increasing $t_V$ further reverses the boundary polarization [Fig.~\ref{fig:2}(c)] and restores the skin-suppressed phase at $t_V = \pi$ [Fig.~\ref{fig:2}(d)]. Away from these points, the LSE reappears with opposite boundary preference, indicating a drive-induced \textit{edge-to-edge communication} through these skin-suppressed points.


In this context, the localization length $\xi$, that governs the relaxation timescales \cite{dissipative_topology2,NH_theory6,NHCreutz1}, accurately captures the dynamical signatures of such transitions. In the static limit, $\xi \propto 2/\ln(|r^2|)$, where $r = \sqrt{ |(t_V - \gamma)/(t_V + \gamma)| }$ \cite{NHCreutz1}. Extending this to the driven case yields effective parameters $t_V^{\mathrm{eff}} = \cos(\gamma)\sin(t_V)\cos(2t_D)$ and $\gamma^{\mathrm{eff}} = \sin(\gamma)$, and helps in defining an effective localization length $\xi_{\mathrm{driven}}$ (see SM). As shown in Fig.~\ref{fig:2}(e), $\xi_{\mathrm{driven}}$ exhibits a periodic dependence on $t_V$, diverging at the skin-suppressed points ($t_V = 2n\pi/T$) and collapsing between two consecutive ones, where exceptional points emerge (at $\xi_{\text{driven}}=0$). The regions between two successive exceptional points correspond to trivial dissipative phases, revealing a rich drive-controlled hierarchy of non-Hermitian localization phenomena. Moreover, the dynamical imprint of such unipolar–bipolar crossover appears as a drive-induced \textit{chiral–helical damping transition}. In the static limit, the damping wavefront is strongly uni-directional, imposing an \textit{information constraint} where excitations propagate efficiently only along one preferred direction \cite{information_constraint1,information_constraint2}. Under periodic driving, however, the emergence of $\mathbb{Z}_2$ points lifts this constraint, converting the one-sided chiral decay into a helical damping pattern (see SM).
\begin{figure}[t]
         \includegraphics[width=\columnwidth]{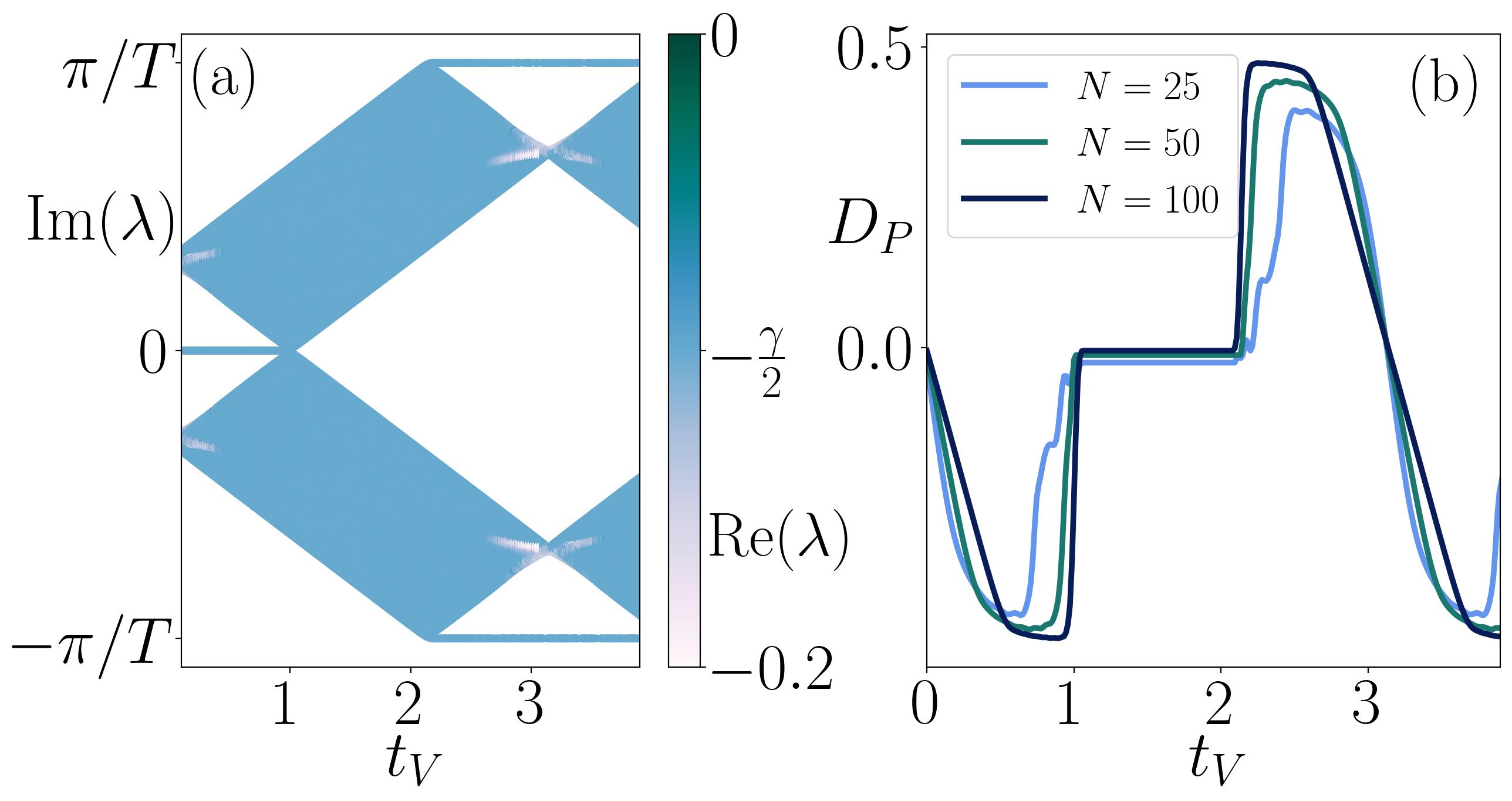}
\caption{(a) Floquet rapidity spectrum versus $t_V$, color‐coded by the real eigenvalues of $X_F$. 
(b) Polarization drift $D_P$ as a function of $t_V$, showing a one-to-one correspondence with the spectral features in panel (a). Here, we have chosen $L=100$, $t_D=t_H=0.5$, $\gamma_l=\gamma_g=0.3$, and $T=2$.
}

\label{fig:3}
\end{figure} 


\par \textit{Dynamical Polarization and isolation of boundary states.-} In order to elucidate the dynamical imprint of the unipolar–bipolar transition in identifying nontrivial phases, we analyze the time evolution of the single-particle correlation matrix, $\mathbf{C(t)}$. Unlike the static case, where the stationary state follows from a continuous Lyapunov equation, the time-periodic Liouvillian demands a stroboscopic treatment. The correlation dynamics is then governed by, \cite{Floquet_Lindblad1,Floquet_Lindblad2},
\begin{equation}
\mathbf{C(t)} = \mathbf{U(t)}\, \mathbf{C(0)}\, \mathbf{U^{T}(t)} - i\, \mathbf{Q(t)}\, \mathbf{U^{T}(t)},
\end{equation}
where $\mathbf{U(t)} = \mathcal{T} \exp\!\left[-\!\int_0^t \mathbf{X}(t')\, dt'\right]$ and $\mathbf{Q(t)}$ encodes the off-diagonal Floquet Livouillian (see SM). The asymptotic Floquet steady state, $\mathbf{C_F}$, satisfies the discrete Lyapunov equation,
$\mathbf{U(T) C_F} - \mathbf{C_F U^{-T}(T)} = i\mathbf{Q(T)}$. A directly measurable consequence of this dynamics appears in the relaxation of local fermionic densities, quantified by $\tilde{n}_x(t) = n_x(t) - n_x(\infty)$ with $n_x(t) = C_{xA,xA}(t) + C_{xB,xB}(t)$. While under periodic boundary conditions (PBC), the site-averaged density, $\tilde{n}(t) = \sqrt{\sum \tilde{n}_x^2/L}$, distinguishes topological and trivial phases through algebraic and exponential decay, respectively. This contrast, however, vanishes under open boundary conditions (OBC), where all modes decay exponentially. Remarkably, periodic driving reinstates topological fingerprints in the transient dynamics of $n(t)$, motivating the necessity of a more sensitive dynamical probe. To this end, we introduce the dynamical polarization, 
\begin{figure}[t]
         \includegraphics[width=\columnwidth]{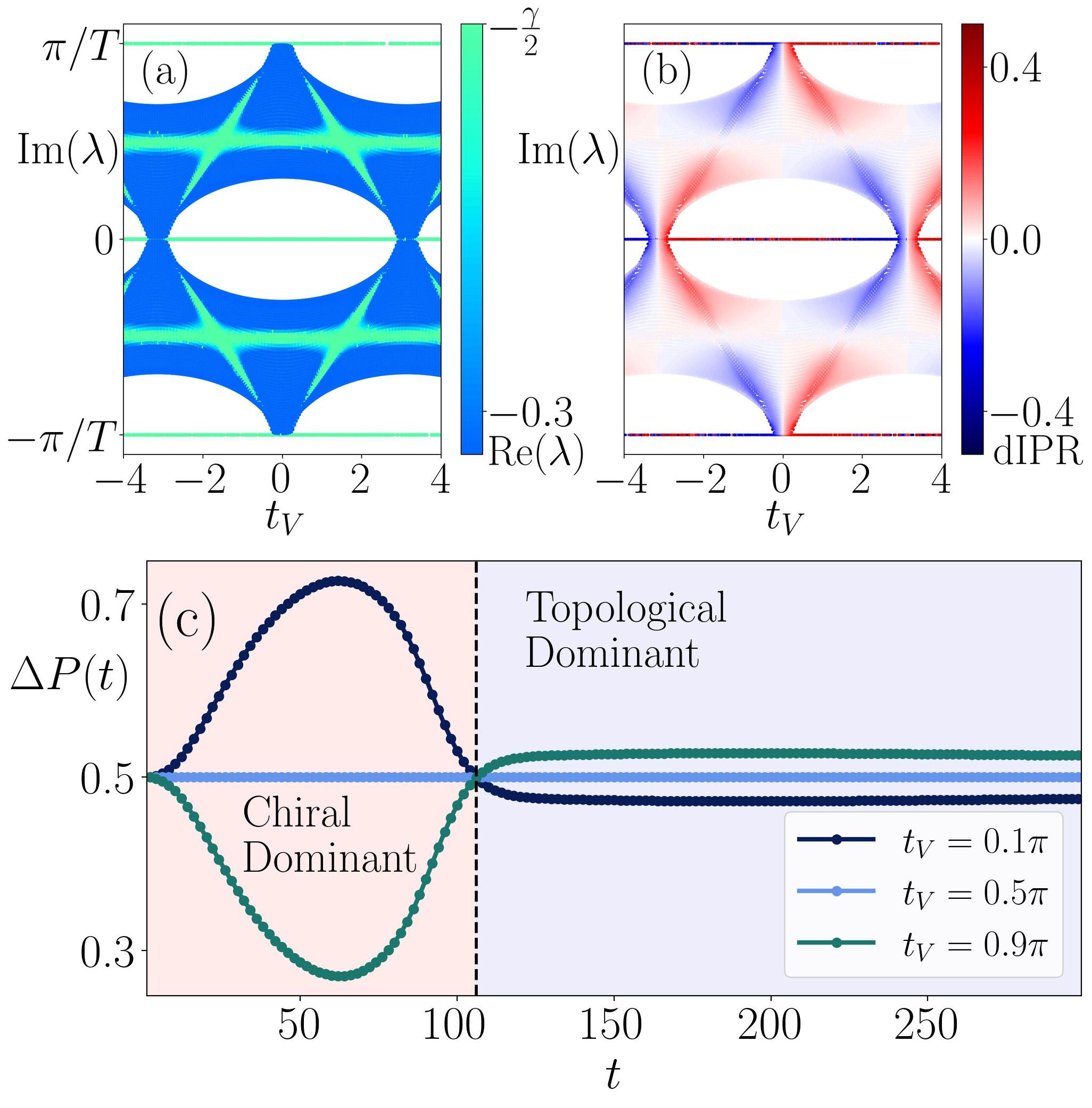}
\caption{(a),(b) Floquet rapidity spectrum as a function of $t_V$, color–coded by the real eigenvalues of $X_F$ and the corresponding dIPR values, respectively. A direct comparison shows that left/right skin modes exhibit enhanced decay rates, whereas topological boundary modes retain the smallest decay rates. 
(c) Time evolution of $\Delta P(t)$, revealing two distinct dynamical regimes: an initial short–time decay governed by a rapidly damped chiral wavefront that depletes the bulk, followed by a long–time regime dominated by robust edge modes. Here, $L=100$, $t_D=1.5$, $t_H=0.5$, $\gamma_l=\gamma_g=0.3$, and $T=2$.
}
\label{fig:4}
\end{figure}
$
\Delta P(t) = \frac{\sum_{j=1}^N j\, n_j(t)}{N \sum_{j=1}^N \, n_j(t)},
$ which measures the center-of-mass motion of the fermionic density and directly tracks chiral damping. Starting from a uniformly filled state with $\Delta P(0) = 1/2$, the evolution of $\Delta P(t)$ reflects asymmetric particle accumulation at the edges (see SM), signaling unidirectional transport. In the static case, $\Delta P(t)$ relaxes over similar timescales for both topological and trivial phases, blurring the topological distinction. Under periodic driving, however, the relaxation occurs on sharply different timescales (more details in the SM), a behavior that directly stems from the drive-renormalized bulk localization length, $\xi_{\text{driven}}$.
Moreover, as shown in Fig. \ref{fig:2}(e), $\xi_{\text{driven}}$ remains significantly smaller than its static counterpart over an extended range of $t_V$, leading to a stronger spatial confinement of the skin modes. As a consequence, we find that, $\Delta P(t)$ remains pinned near its extremal values (0 or 1) for extended periods in the topological phase compared to the static case, whereas it decays rapidly in the trivial one. To quantitatively distinguish these regimes, we introduce the \textit{polarization drift} defined by,
\begin{equation}
    D_P = \Big[\,\Delta P(t,t_V) - \Delta P(t,t_V = (2n+1)\pi/T)\,\Big]_{t > t_{\text{max}}}
    \label{eq:drift}
\end{equation}
where $t_{\text{max}}$ denotes the time at which $\Delta P(t)$ attains its extremal value, and $t$ is chosen to be larger than $t_{\text{max}}$ yet remains within the transient window. As shown in Fig.~\ref{fig:3}(a), the rapidity spectrum of the Floquet Liouvillian $\mathbf{X_F}$ (color-mapped by its real component, indicating decay rates) identifies the trivial regime, where $\xi_{\text{driven}}$ [Fig.~\ref{fig:2}(e)] remains nearly flat and comparable to its value at $t_V = (2n+1)\pi/T$ (e.g., $t_V = \pi/2$), leading to a bidirectional $\mathbb{Z}_2$ skin effect and steady polarization near 0.5. In contrast, in the topological regime $\Delta P(t)$ stays locked at 0 or 1 over long times, yielding a finite $D_P$, whereas in the trivial phase it saturates rapidly to the reference value, giving $D_P = 0$. As shown in Fig.~\ref{fig:2}(b), the behavior of $D_P$ provides a robust dynamical order parameter: a finite value signals the onset of a topological phase, while a vanishing $D_P$ indicates trivial behavior. This classification is corroborated by the Liouvillian rapidity spectrum, which exhibits increasingly sharp phase boundaries for larger system sizes [Fig.~\ref{fig:3}(b)].
\begin{figure}[t]
         \includegraphics[width=\columnwidth]{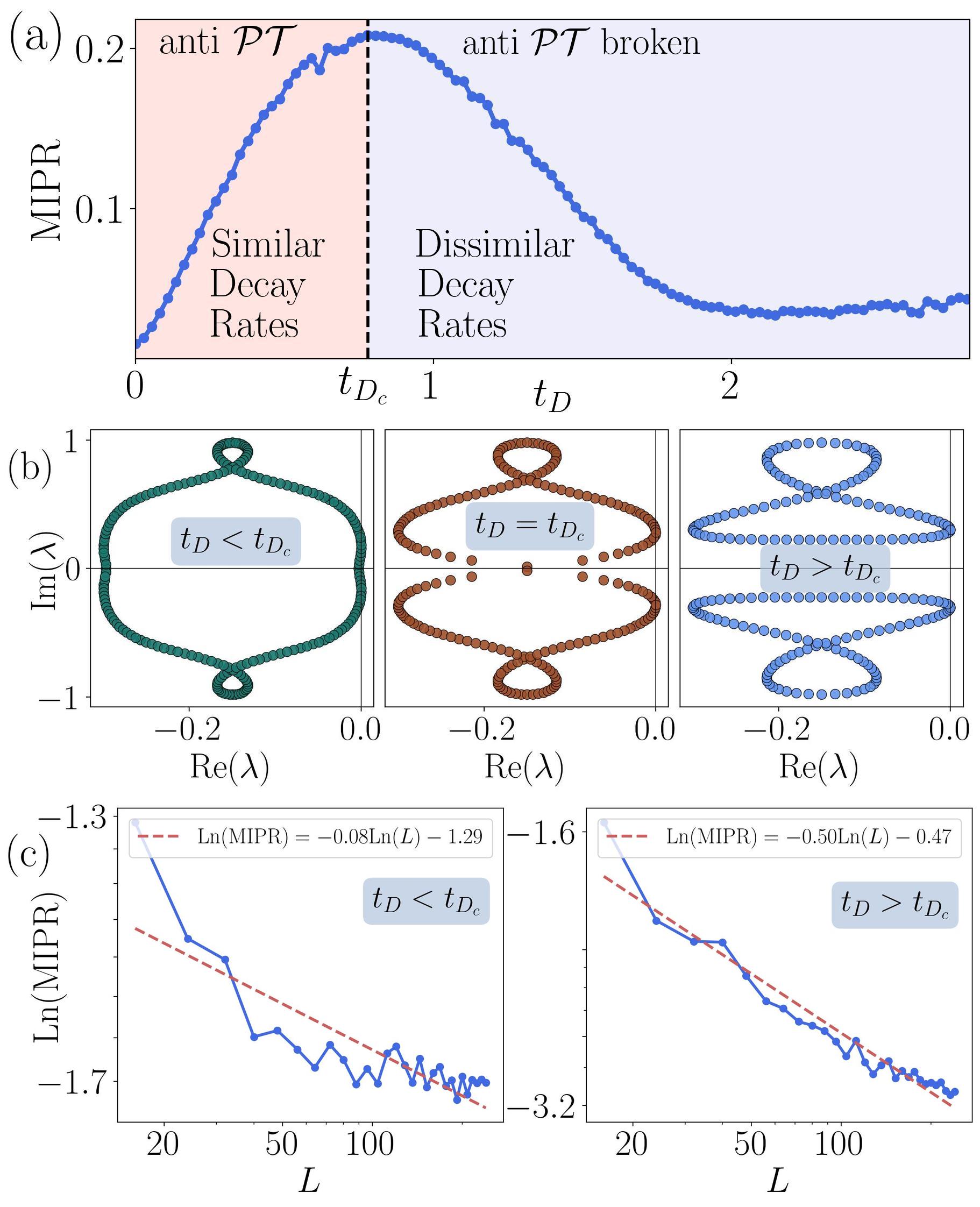}
\caption{(a) MIPR as a function of $t_D$, revealing two distinct regimes: an anti–$\mathcal{PT}$–symmetric phase for $t_D < t_{D_c}$, where all states are skin–localized with nearly identical decay rates, and an anti–$\mathcal{PT}$–broken phase for $t_D > t_{D_c}$, where a subset of states becomes delocalized and acquires decay rates distinct from the skin modes. 
(b) Evolution of the PBC spectrum from point-gap to line-gap topology across $t_{D_c}$, signaling the breakdown of LSE. 
(c) Finite–size scaling of MIPR, showing system-size–independent localization for $t_D < t_{D_c}$ and a growing localization length for $t_D > t_{D_c}$. Here, $t_D=0.5$ ($t_D < t_{D_c}$) and $t_D=1.5$ ($t_D > t_{D_c}$), $t_H=0.5$, $\gamma_l=\gamma_g=0.3$, and $T=2$.
}
\label{fig:5}
\end{figure}
\par While the transient dynamics captured by the drive-renormalized localization parameter successfully differentiate the underlying topological regimes, the boundary modes remain intertwined with the skin-localized bulk states. A genuine isolation of edge states demands distinct decay rates for bulk and boundary excitations. However, as shown in Fig. \ref{fig:3}(a), under weak dissipation both exhibit nearly identical decay rates, owing to uniform real eigenvalues of the damping matrix that enforces a global exponential decay of $\tilde{n}(t)$ across both topological and trivial phases under OBC.
Nevertheless, in the strong dissipation limit, the static model can, in principle, generate nonuniform decay rates, but only at the expense of breaking antiunitary symmetry and suppressing the topological features. Moreover, such variation cannot be systematically controlled. Periodic driving, in contrast, provides a controllable route to selectively modify decay rates without compromising topological protection. To investigate this aspect, we consider a regime where the driving dominates dynamics of the system. In particular, we choose a moderately large value of $t_D$ (specifically, $t_D =1.5$), so that higher winding numbers emerge and multiple edge modes stabilize a globally topological phase \cite{Roy_Floquet_Creutz1}. The rationale behind this choice will become evident shortly. The rapidity spectrum [Fig. \ref{fig:4}(a)] exhibits a clear hierarchy of decay rates, ranging from $-\gamma/2$ (green) to substantially larger magnitudes (blue). Strikingly, these decay rates correlate with distinct localization properties. Although all states should exhibit exponential localization under LSE, the drive induces an unconventional redistribution. Remarkably, we find out that, during the drive-induced unipolar–bipolar transition, \textit{not} all the states switch their localization \textit{abruptly}. Instead, a fraction of states gradually delocalizes across the lattice before relocalizing on the opposite edge, forming a mixed regime where extended and skin-localized modes coexist. We identify this as an \textit{intermediate skin phase} mediating as a bridge between unipolar and bipolar regimes. To decipher this asymmetry, we compute the directional inverse participation ratio (dIPR), defined as,
\begin{equation}
\text{dIPR}(\psi_n) = \text{sgn}\left[\sum_{j=1}^L (j - L/2 - \delta) |\psi_{n,j}|\right] \, \text{IPR}(\psi_n),
\end{equation}
with $\delta \in (0,0.5)$. Positive (negative) dIPR denotes right (left) localization, while values near zero indicate delocalized states. Fig. \ref{fig:4}(b) presents the Liouvillian spectrum color-coded by its dIPR values. A direct comparison with Fig. \ref{fig:4}(a) reveals that at certain values of $t_V$ the states with the fastest decay rates (blue in Fig. \ref{fig:4}a) correspond to skin-localized bulk modes (non-zero dIPR in Fig. \ref{fig:4}b), while the most long-lived states with the smallest decay rates are either extended (dIPR $\sim$ 0) or correspond to the robust topological edge modes. This establishes that periodic driving dynamically discerns the lifetimes of topological boundaries from the dissipative bulk.
\par The dynamical consequence of this decoupling is also reflected in the temporal evolution of polarization. As illustrated in Fig. \ref{fig:4}(c), for $t_V = 0.1\pi$ and $0.9\pi$, $\Delta P(t)$ initially approaches 0 or 1 depending on the localization direction. However, the skin modes characterized by extremal damping rates decay rapidly before reaching full saturation, eventually stabilizing around $\Delta P(t) \sim 0.5$, where only the long-lived edge modes at the two boundaries persist. Consequently, the temporal evolution naturally separates into two regimes: a short-time regime dominated by a rapidly decaying chiral wavefront that empties the skin-localized bulk, followed by a long-time regime governed by resilient topological edge states.
\par Further, to achieve a more transparent characterization of the decay-rate spectrum, we identify $t_D$ as a controllable knob, clarifying why a stronger $t_D$ is required to induce state-selective decay. Varying $t_D$ drives the system through an intermediate skin phase with coexisting localized and extended modes. The emergence of such coexistence suggests the presence of a critical threshold, $t_{D_c}$, that signals an abrupt change in the localization behavior, reminiscent of the critical skin effect \cite{critical_skin}. To characterize this crossover, we compute the mean IPR, $\text{MIPR} = \frac{1}{L} \sum_n \text{IPR}(\psi_n)$, which provides a global measure of localization. As shown in Fig. \ref{fig:5}(a), the MIPR varies non-monotonically with $t_D$, showing a distinct critical point ($t_{D_c}$) where it rapidly decreases on either side of it. For, $t_D < t_{D_c}$, the spectrum is dominated by skin-localized modes, whereas $t_D > t_{D_c}$ marks the onset of the intermediate skin regime. This transition coincides with the spontaneous breaking of anti-$\mathcal{PT}$ symmetry. While the 
$t_D < t_{D_c}$ regime remains anti-$\mathcal{PT}$ symmetric with distinct PBC and OBC spectra, for 
$t_D > t_{D_c}$, the anti-$\mathcal{PT}$ symmetry is broken, leading to a qualitative change in the spectral topology. As illustrated in Fig. \ref{fig:5}(b), the point-gap topology corresponding to PBC closes across the transition and evolves into a line-gap structure at higher $t_D$. For the representative value $t_D=1.5$ (which we used in Fig. \ref{fig:4}), the system lies deep in this symmetry-broken regime, where the spectrum exhibits complex topology and a hierarchy of decay rates. Moreover, as seen in Fig. \ref{fig:4}(b), localized states tend to cluster near the topological edges, either around $\lambda=0$ or $\pm \pi/T$ while the extended bulk states occupy the intervening regions, leading to an energy-dependent separation between localized and extended states, which we refer to as an \textit{anomalous mobility edge.} Furthermore, the scaling behavior of localization also changes qualitatively across such transition. in the symmetry-preserving phase, the MIPR remains size-independent, while in the broken phase it follows a logarithmic scaling, $\text{ln(MIPR)} = -\alpha \text{ln}(N) - C$. Numerical fits shown in Figs. \ref{fig:5}(c),(d) yield $\alpha \approx 0$ in the unbroken phase and $\alpha \approx 0.5$ in the broken phase, implying that some of the states are localized, while the remainder remain extended. This suggests the emergence of a \textit{scale-free localization} behavior in the anti-$\mathcal{PT}$ broken regime, where the localization length scales linearly with system size. Consequently, this signals the onset of delocalized states that exhibit decay rates distinct from those of the skin modes.  
\par \textit{Conclusion.-} 
Overall, our results provide a coherent framework for diagnosing nontrivial phases. While a drift in the transient polarization reliably captures the underlying topology, tuning the drive-induced parameters enables a scale-free localization regime, which naturally enables state-selective decay, allowing protected boundary modes as the sole long-lived excitations. Altogether, our results demonstrate a compact and physically relevant route for diagnosing and manipulating NH topological phases through their dynamical responses.
\par \textit{Acknowledgements.-} KR and SS sincerely acknowledge Koustabh Gogoi and Latu Kalita for fruitful discussions.  Also, KR acknowledges the research fellowship from the MoE, Government of India.



\bibliography{refs}

\clearpage
\pagebreak
\widetext
\setcounter{equation}{0}
\setcounter{figure}{0}
\setcounter{table}{0}

\renewcommand{\theequation}{S\arabic{equation}}
\renewcommand{\thefigure}{S\arabic{figure}}
\renewcommand{\thetable}{S\arabic{table}}
\renewcommand{\bibnumfmt}[1]{[S#1]}
\renewcommand{\citenumfont}[1]{S#1}
\newcommand{\bk}{\boldsymbol\kappa}

\newcommand{\SI}{Supplementary Material}
\newcommand{\beginsupplement}{%
  \setcounter{equation}{0}
  \renewcommand{\theequation}{S\arabic{equation}}%
  \setcounter{table}{0}
  \renewcommand{\thetable}{S\arabic{table}}%
  \setcounter{figure}{0}
  \renewcommand{\thefigure}{S\arabic{figure}}%
  \setcounter{section}{0}
  \renewcommand{\thesection}{S\Roman{section}}%
  \setcounter{subsection}{0}
  \renewcommand{\thesubsection}{S\Roman{section}.\Alph{subsection}}%
}

\begin{center}
\textbf{\large Supplemental Materials For: Controlling Dissipative Topology Through Floquet Driving: From Transient Diagnostics to Boundary States Isolation }
\end{center}

\makeatletter
\def\l@section#1#2{%
  \@dottedtocline{1}{1.5em}{2.3em}{#1}{#2}%
}
\def\l@subsection#1#2{%
  \@dottedtocline{2}{3.0em}{3.2em}{#1}{#2}%
}
\makeatletter
\tableofcontents

\section*{S1: Third Quantization}
Here, we provide a full derivation for the third quantization technique used in the main text. The subsequent discussion follows the adjoint-fermion approach used in the literature \cite{third_quantization1,third_quantization2,third_quantization3}.
We start by rewriting all (the complex) fermions in terms of real Majorana operators. For each lattice site $j$ and sublattice $\eta$ (with $\eta\!=\!A,B$), we define
\begin{equation}
    c_{j,\eta}=\frac{w_{2j-1,\eta}- i w_{2j,\eta}}{2}\,,\qquad
    c^\dagger_{j,\eta}=\frac{w_{2j-1,\eta}+ i w_{2j,\eta}}{2}\,,
\label{eq:ctow}
\end{equation}
where $w_{a,\eta}$ are the real Majorana operators satisfying
\begin{equation}
    \{w_{a,\eta},w_{b,\beta}\}=2\delta_{a,b}\,\delta_{\eta,\beta}\, ,
\end{equation}
We collect all Majorana operators into a single column vector
\[
\underline{w} = \big\{ w_{1,A},w_{1,B},w_{2,A},w_{2,B},\dots \big\}^{T},
\]
whose length is $4N$ for a chain with $N$ lattice sites and two internal degrees. In such a Majorana basis, the Hamiltonian and any linear jump operator take the quadratic and linear forms
\begin{align}
    H &= \sum_{a,b=1}^{4N} w_a \, \mathbf{H}_{a b}\, w_b, \label{eq:HMdef}\\
    L_\alpha &= \sum_{p=1}^{4N} l^{(\alpha)}_{p}\, w_p \, ,
\end{align}
where $\mathbf{H}$ is a real antisymmetric $4N\times 4N$ matrix and each $l^\alpha$ is a column vector of Majorana coefficients. The construction of $\mathbf{H}_{ab}$ and $l_p^{(\alpha)}$ for the Creutz Ladder is done by inserting the Majorana decomposition \ref{eq:ctow} into the second quantized Hamiltonian and collecting coefficients. Now, the Lindblad master equation in the superoperator form is written as,
\small{
\begin{equation}
\frac{d\rho}{dt} = \mathcal{L}\rho(t) = -i[H, \rho] + \sum_{\alpha} \left( L_\alpha \rho L_\alpha^{\dagger} - \frac{1}{2} \left\{ L_\alpha^{\dagger} L_\alpha, \rho \right\} \right).
\label{eq:lindblad}
\end{equation}
}
where $\mathcal{L}$ is the superoperator acting on the density matrix $\rho$ consisting of the unitary part $\mathcal{L}_0[\rho]$ and the dissipative part $\mathcal{D}[\rho]$.
To handle the full dynamics including the quantum jumps, we move to the Liouville-Fock space $\mathcal{K}$, i.e., the Fock space of operators. In $\mathcal{K}$ we represent basis elements by products of Majorana operators,
\begin{equation}
    P_\alpha = \prod_{i=1}^{4N} w_i^{\alpha_a},\qquad \alpha_i\in\{0,1\},
\end{equation}
The inner product is given by as $\langle{x} {y} \rangle=2^{-4N} {\rm tr} \left[x^\dagger y \right]$ and the adjoint (or ``super'') fermion annihilation and creation operators $\phi_a,\phi_a^\dagger$ acting on $\ket{P_\alpha}\in\mathcal{K}$ via
\begin{equation}
    \phi_i \ket{P_\alpha} = \delta_{\alpha_i,1}\, \ket{w_i P_\alpha},\qquad
    \phi_i^\dagger \ket{P_\alpha} = \delta_{\alpha_i,0}\, \ket{w_i P_\alpha}.
\end{equation}
These obey the fermion canonical anti-commutation relations
\begin{equation}
    \{\phi_i,\phi_j^\dagger\}=\delta_{i j}.
\end{equation}
The action of the Majorana operators $w_a$ is given by,
\begin{align}
    \ket{w_i P_{\alpha}} &= \left( \phi_i^\dagger + \phi_i \right) \ket{P_{{\alpha}}}, \nonumber\\[4pt]
    \ket{P_{{\alpha}} w_i} &= \mathcal{P} \left( \phi_i^\dagger - \phi_i \right) \ket{P_{{\alpha}}},
    \label{idnty1}
\end{align}
where $\mathcal{P}$ is the fermion parity operator defined as 
    $\mathcal{P} = (-1)^{\sum_a \alpha_a} 
    = \exp\!\left( i\pi \mathcal{N} \right)$, $\mathcal{N}$ being the number operator. Using the anticommutation relations, it can be shown that the density matrix $\rho(t)$ can now be represented by the basis element $\ket{P_{\alpha}}$. The unitary part of the Liouvillian reads as
\begin{align}
    \mathcal{L}_0[\rho] 
    &= -i [ H, \rho(t) ] \nonumber\\[2pt]
    &= -i\!\sum_{a,b} \mathbf{H}_{a b} 
       \!\left( \ket{w_a w_b P_{\alpha}} - \ket{P_{\alpha} w_a w_b} \right) \nonumber\\[2pt]
    &= -4i \!\sum_{a,b} \phi_a^\dagger \mathbf{H}_{a b} \phi_b 
       \quad \nonumber\\[2pt]
    &= -4i\,\boldsymbol{\phi}^\dagger \!\cdot\! \mathbf{H}_M\,\boldsymbol{\phi}\, .
\end{align}

The non-unitary part can be written as
{\small
\begin{align}
    \mathcal{D}[\rho] 
    &= -\tfrac{1}{2}\!\sum_{n} 
       \!\left( \{L_n^\dagger L_n, \rho(t)\} - 2 L_n^\dagger \rho(t) L_n \right) \nonumber\\[2pt]
    &= -\tfrac{1}{2}\!\sum_{n,i,j} l_{n,i} l_{n,j}^*
       \!\left( \ket{w_i w_j P_{\alpha}} + \ket{P_{\alpha} w_i w_j} 
       - 2 \ket{w_i P_{\alpha} w_j} \right) \nonumber\\[2pt]
    &= \sum_{n,i,j} \tfrac{(1+\mathcal{P})}{2} 
       \!\Big[ \phi_i^\dagger (l_{n,i} l_{n,j}^* - l_{n,j} l_{n,i}^*) \phi_j^\dagger 
       - \phi_i^\dagger (l_{n,i} l_{n,j}^* + l_{n,j} l_{n,i}^*) \phi_j \Big] \nonumber\\[2pt]
    &\quad + \sum_{n,i,j} \tfrac{(1-\mathcal{P})}{2} 
       \!\Big[ \phi_i (l_{n,i} l_{n,j}^* - l_{n,j} l_{n,i}^*) \phi_j 
       - \phi_i (l_{n,i} l_{n,j}^* + l_{n,j} l_{n,i}^*) \phi_j^\dagger \Big] \nonumber\\[2pt]
    &= \sum_{i,j} \tfrac{(1+\mathcal{P})}{2} 
       \!\Big[ \phi_i^\dagger (\mathcal{M}_{i j} - \mathcal{M}_{j i}) \phi_j^\dagger 
       - \phi_i^\dagger (\mathcal{M}_{i j} + \mathcal{M}_{j i}) \phi_j \Big] \nonumber\\[2pt]
    &\quad + \sum_{i,j} \tfrac{(1-\mathcal{P})}{2} 
       \!\Big[ \phi_i (\mathcal{M}_{i j} - \mathcal{M}_{j i}) \phi_j 
       - \phi_i (\mathcal{M}_{i j} + \mathcal{M}_{j i}) \phi_j^\dagger \Big],
    \label{eq:dissipator}
\end{align}
}

where $\mathcal{M}_{a b} = \sum_m l_{m,a} l_{m,b}^*$. Thus, the full Liouvillian reads as
\begin{align}
    \mathcal{L} 
    &= -4i\,\boldsymbol{\phi}^\dagger \!\cdot\! \mathbf{H}_M\,\boldsymbol{\phi} \nonumber\\[2pt]
    &\quad + \tfrac{1+\mathcal{P}}{2} 
       \!\Big[ \boldsymbol{\phi}^\dagger \!\cdot\! (\mathbf{M - M^T}) \boldsymbol{\phi}^\dagger
       - \boldsymbol{\phi}^\dagger \!\cdot\! (\mathbf{M + M^T}) \boldsymbol{\phi} \Big] \nonumber\\[2pt]
    &\quad + \tfrac{1-\mathcal{P}}{2} 
       \!\Big[ \boldsymbol{\phi} \!\cdot\! (\mathbf{M - M^T}) \boldsymbol{\phi}
       - \boldsymbol{\phi} \!\cdot\! (\mathbf{M + M^T}) \boldsymbol{\phi}^\dagger \Big].
    \label{fullLiouvillian}
\end{align}

Finally, for the physical sector with an even number of fermions, we set $\mathcal{P}=+1$, so that Eq.~\eqref{fullLiouvillian} reduces to the even-parity Liouvillian $\mathcal{L}_+$ used in the main text, completing the derivation.

\begin{align}
	\mathcal{L}_+= \frac{1} {2}  
	\begin{pmatrix} 
		\underline{\phi}^\dagger \cdot &  \underline{\phi} \cdot
	\end{pmatrix} 
	\begin{pmatrix} 
		-\mathbf{X}^\dagger &  -i \mathbf{Y} \\
		0 & \mathbf{X}
	\end{pmatrix} 
	\begin{pmatrix} 
		\underline{\phi}^\dagger \\  \underline{\phi}
	\end{pmatrix}  -{A_0} \ ,
 \label{Liouvillianplus}
\end{align}
where we define \textbf{}$X=-4 i\mathbf{H}_{\rm M} +  2(\mathbf{M} +\mathbf{M}^T)$, $Y=2i(\mathbf{M}-\mathbf{M}^T)$, and $A_0=\frac{1}{2} {\rm Tr} [\mathbf{X}]$. Here, $\underline{\phi}^\dagger$ and $\underline{\phi}$ represent the vector form of the adjoint creation and annihilation operators, respectively. 
Since $\mathcal{L}_+$ takes an upper triangular form, the spectrum of the Liouvillian, which is also called the Lindblad spectrum, is completely determined by the eigenvalues of $\mathbf{X}$.

\section*{S2: Floquet-Lindblad framework}
\label{sec:floquet}

In order to understand the effects of dissipation in a periodically driven bulk, we first generalize the Floquet formalism to the case of Lindbladian time evolution. The dynamics of the density matrix $\hat{\rho}$ is governed by the time-dependent master equation
\begin{equation}
    \frac{d\hat{\rho}}{dt} = \mathcal{L}(t)[\hat{\rho}],
\end{equation}
where $\mathcal{L}(t)$ denotes the time-dependent Liouvillian superoperator. Over one period of the drive $T$, the stroboscopic evolution is captured by the propagator
\begin{equation}
    \hat{U}(t_0 + T, t_0) = \hat{T}\, \exp\!\left[\int_{t_0}^{t_0+T} dt\, \mathcal{L}(t)\right],
\end{equation}
where $\hat{T}$ denotes time ordering. Analogous to unitary Floquet theory, we define an effective \textit{Floquet Liouvillian} $\mathcal{L}_F(t_0)$, such that
\begin{equation}
    \hat{U}(t_0 + T, t_0) = e^{\mathcal{L}_F(t_0) T}.
\end{equation}

To construct $\mathcal{L}_F(t_0)$ explicitly, we consider a piecewise-constant drive as used in the main text, where the system evolves under two distinct Liouvillians $\mathcal{L}_1$ and $\mathcal{L}_2$ for two equal half-periods $T/2$. Using the third quantized framework introduced in the previous section, we write the total evolution over one period as
\begin{equation}
    e^{\mathbf{{A}_F} T} = e^{\mathbf{A_2} T/2} e^{\mathbf{A_1} T/2},
\label{eq:lyap_define}
\end{equation}
where $\mathbf{A_i}$ are the shape-matrices of the respective Liouvillian in the third-quantized space:
\begin{equation}
\mathbf{A_i} =
\begin{pmatrix}
    -\mathbf{X_i}^\dagger & i\mathbf{Y_i} \\
    0 & \mathbf{X_i}
\end{pmatrix},
\qquad i = 1,2.
\end{equation}
The exponential of $A_i$ can be evaluated using its block structure, namely,
\begin{equation}
e^{\mathbf{A_i} T/2} =
\begin{pmatrix}
    e^{-\mathbf{X_i}^\dagger T/2} & i\mathbf{Z_i} \\
    0 & e^{\mathbf{X_i} T/2}
\end{pmatrix}.
\end{equation}
Here, the off-diagonal block $\mathbf{Z_i}$ can be determined by requiring that the exponential commutes with $A_i$,
\begin{equation}
[\mathbf{A_i}, e^{\mathbf{A_i} T/2}] = 0.
\end{equation}
Evaluating this condition gives a Lyapunov-type matrix equation for each driving segment, such as
\begin{equation}
\mathbf{X_i}^\dagger \mathbf{Z_i} + \mathbf{Z_i} \mathbf{X_i} = e^{-\mathbf{X_i}^\dagger T/2} \mathbf{Y_i} - \mathbf{Y_i} e^{\mathbf{X_i} T/2},
\qquad i = 1,2.
\end{equation}
These two Lyapunov equations fully determine $\mathbf{Z_1}$ and $\mathbf{Z_2}$ for the first and second half-periods, respectively. Finally, the effective Floquet-Lindbladian over one full driving period takes the block form
\begin{equation}
e^{\mathbf{A_F} T} =
\begin{pmatrix}
    e^{-\mathbf{X_F}^\dagger T} & -i\mathbf{Q} \\
    0 & e^{\mathbf{X_F} T}
\end{pmatrix}
\label{eq:AF}
\end{equation}
where the matrices $\mathbf{X_F}$ and $\mathbf{Q}$ are obtained by combining the two half-period propagators $e^{\mathbf{A_1} T/2}$ and $e^{\mathbf{A_2} T/2}$ using Eq.~\eqref{eq:lyap_define}. The resulting effective generator $\mathcal{L}_F$ encapsulates the full stroboscopic evolution.
\noindent
To derive the evolution of the covariance matrix, we define
\[
C_{j,k}(t) = \mathrm{Tr}\!\left[w_j w_k \rho(t)\right] - \delta_{j,k},
\]
whose time dependence follows from the equation,
\begin{equation}
\frac{d}{dt}\mathbf{C}(t)
= -\mathbf{X}^T \mathbf{C}(t)
  - \mathbf{C}(t)\mathbf{X}
  + i\mathbf{Y}.
\label{eq:cov_evol}
\end{equation}

\noindent
The corresponding dynamics can be expressed more compactly as
\begin{equation}
\frac{d}{dt}\mathbf{D}(t)
= [\mathbf{A}(t), \mathbf{D}(t)],
\qquad
\mathbf{D}(t) \equiv
\begin{pmatrix}
\mathbf{1} & \mathbf{C}(t) \\
\mathbf{0} & \mathbf{0}
\end{pmatrix},
\label{eq:D_dynamics}
\end{equation}
which is satisfied by the formal solution
\begin{equation}
\mathbf{D}(t) =
\mathcal{T}
\exp\!\left( \int_{0}^{t} d\tau\, \mathbf{A}(\tau) \right)
\mathbf{D}(0)\,
\tilde{\mathcal{T}}
\exp\!\left( -\int_{0}^{t} d\tau\, \mathbf{A}(\tau) \right),
\label{eq:D_formal_solution}
\end{equation}
where $\mathcal{T}$ and $\tilde{\mathcal{T}}$ denote time- and anti–time-ordering, respectively.
The Floquet steady state corresponds to a fixed point of this evolution, i.e., a state for which $\mathbf{D}(t)$ remains unchanged after one full driving period $T$. This condition yields
\begin{equation}
\mathbf{D}_F = e^{\mathbf{A}_F T}\, \mathbf{D}_F\, e^{-\mathbf{A}_F T}
\quad \Rightarrow \quad
\mathbf{C}_F e^{\mathbf{X}_F^T}
- e^{-\mathbf{X}_F^{\dagger T}}\mathbf{C}_F
= i\mathbf{Q},
\label{eq:discrete_lyapunov}
\end{equation}
where Eq.~\eqref{eq:AF} was used to obtain a \textit{discrete-time Lyapunov equation} 
for the Floquet steady-state covariance matrix $\mathbf{C}_F$.
\section*{S3: Floquet GBZ and bulk localization length}

In this section, we outline the detailed procedure for constructing the Floquet generalized Brillouin zone (FGBZ) using symmetric time frames and explain how this approach yields an analytic version of the driven analogue of the bulk localization parameter. For static non-Hermitian systems, the GBZ is obtained from the characteristic equation
\begin{equation}
    \det[X(\beta)-E]=0,
\end{equation}
which produces a polynomial of degree $2N$ in $\beta$
~\cite{GBZ1,WangGBZ}. In Floquet systems, however, the stroboscopic evolution typically generates effective long-range couplings, resulting in characteristic equations involving infinite-order polynomials in $\beta$ (where $\beta=e^{ik}$), which makes a direct FGBZ construction significantly more difficult.

A major simplification emerges upon working in \emph{symmetric time frames}, where the Floquet Liouvillian takes the form of a massless Dirac operator. By identifying $t=T/2$ as the time-reversal-symmetric point, the evolution over one period can be divided into two segments, $F$ and $G$, which are chiral-symmetric partners. This leads to the symmetric-frame propagators
\begin{equation}
    U_{\mathcal{L},1}=FG, \qquad 
    U_{\mathcal{L},2}=GF.
\end{equation}
In each frame, the effective damping matrix becomes a product of three exponentials; for example,
\begin{equation}
    X_{F,1}
    =e^{-i d_x(k)\sigma_x T/4}\,
     e^{-i d_z(k)\sigma_z T/2}\,
     e^{-i d_x(k)\sigma_x T/4},
\end{equation}
and an analogous expression holds for $X_{F,2}$. Using Euler identities for Pauli matrices, one may write these products in a compact form as
\begin{equation}
    X_{j,k}
    = n_0(k) + i\,\mathbf{n}_j(k)\cdot\boldsymbol{\sigma},
    \qquad j=1,2,
\end{equation}
where $n_0(k)=\cos[\lambda(k)]$ and $\mathbf{n}_j(k)$ is constructed from suitable trigonometric combinations of the underlying static vectors. For the step-driven protocol, these reduce to the effective $d$-vectors of the form
\begin{subequations}
\begin{align}
d_{1x} &= \sin(d_x)\cos(d_z), & d_{1z} &= \sin(d_z), \\
d_{2x} &= \sin(d_x), & d_{2z} &= \cos(d_x)\sin(d_z),
\end{align}
\label{eq:d-vectors}
\end{subequations}
with
\begin{equation}
d_x = t_V + 2t_D\cos k, \qquad 
d_z = 2t_H\sin k + i\gamma/2.
\end{equation}
Despite this concise structure, these expressions encode long-range interactions via functions for example $\sin[F(\cos k)]$, implying that the characteristic equation may contain infinitely many powers of $\beta$. To render the problem tractable, we perform a controlled Taylor expansion in the hopping amplitudes (corresponding to $t_H=t_D$), assuming they are small enough that the expansion can be truncated at a finite order $n$. This approximation reduces the characteristic equation to a quartic polynomial in $\beta$ given as,
\begin{equation}
    \lambda^2(\beta)
    = d_{j,x}^2(\beta) + d_{j,z}^2(\beta),
    \qquad j=1,2,
\end{equation}
which, following a second-order Taylor expansion, assumes a general form,
\begin{equation}
    \lambda^2(\beta)
    = \sum_{j=-2}^{2} D^{j}\,\beta^{j}.
\end{equation}
Now consider two solutions $\beta$ and $\beta'(=\beta e^{i\theta})$ of equal magnitude. Subtracting the corresponding characteristic equations yields
\begin{equation}
    0=\sum_{j=-2}^{2} D^{j}\beta^{j}\big(1-e^{ij\theta}\big),
\end{equation}
from which one extracts all allowed $\beta$ over $\theta\in[0,2\pi]$. Sorting the solutions by their magnitude and selecting those satisfying $|\beta_{2N}|=|\beta_{3N}|$ (with $N$ sampling points in $\theta$) yields the Floquet GBZ, denoted $D_\beta$, as shown in the inset of Fig.~\ref{fig:1} of the main text.

At this stage, it is important to note that the FGBZ directly determines the bulk localization parameter, since the radius $r=|\beta|$ gives the localization length $\xi\propto 2/\ln(r^2)$. For static systems, this reduces to the well-known expression \cite{NHCreutz1},
\begin{equation}
r=\sqrt{\left|\frac{t_V-\gamma}{\,t_V+\gamma\,}\right|}.
\end{equation}
Similarly, in the driven case, an analogous expression emerges once we extract the drive-induced effective parameters $t_V^{\mathrm{eff}}$ and $\gamma^{\mathrm{eff}}$. This requires evaluating $d_x(k)$ and $d_z(k)$ at the symmetric points $k=0$ and $\pi$ which yields
\begin{align}
d_x(0)=t_V+2t_D, \qquad & d_z(0)=\gamma,\\
d_x(\pi)=t_V-2t_D, \qquad & d_z(\pi)=\gamma.
\end{align}
This leads to the effective quantities
\begin{align}
d_x^{\mathrm{eff}}(0)&=\sin(t_V+2t_D)\cos\gamma,\\
d_x^{\mathrm{eff}}(\pi)&=\sin(t_V-2t_D)\cos\gamma,
\end{align}
which we now identify in terms of drive-induced parameters,
\begin{align}
d_x^{\mathrm{eff}}(0)&=t_V^{\mathrm{eff}}+2t_D^{\mathrm{eff}},\\
d_x^{\mathrm{eff}}(\pi)&=t_V^{\mathrm{eff}}-2t_D^{\mathrm{eff}}.
\end{align}
Solving these results in the drive-induced parameters as,
\begin{equation}
t_V^{\mathrm{eff}}
=\cos\gamma\,\sin(t_V)\cos(2t_D), \qquad \text{and} \qquad \gamma^{\mathrm{eff}}=\sin\gamma.
\end{equation}
Using these effective parameters, we obtain a schematic analytic form of the driven bulk localization length $\xi_{\mathrm{driven}}$, which exhibits periodic behavior and diverges at the skin-suppressed points where the Floquet GBZ collapses to a unit circle (see Fig.~\ref{fig:2}(e) of the main text).
\begin{figure}[t]
         \includegraphics[width=0.7\columnwidth]{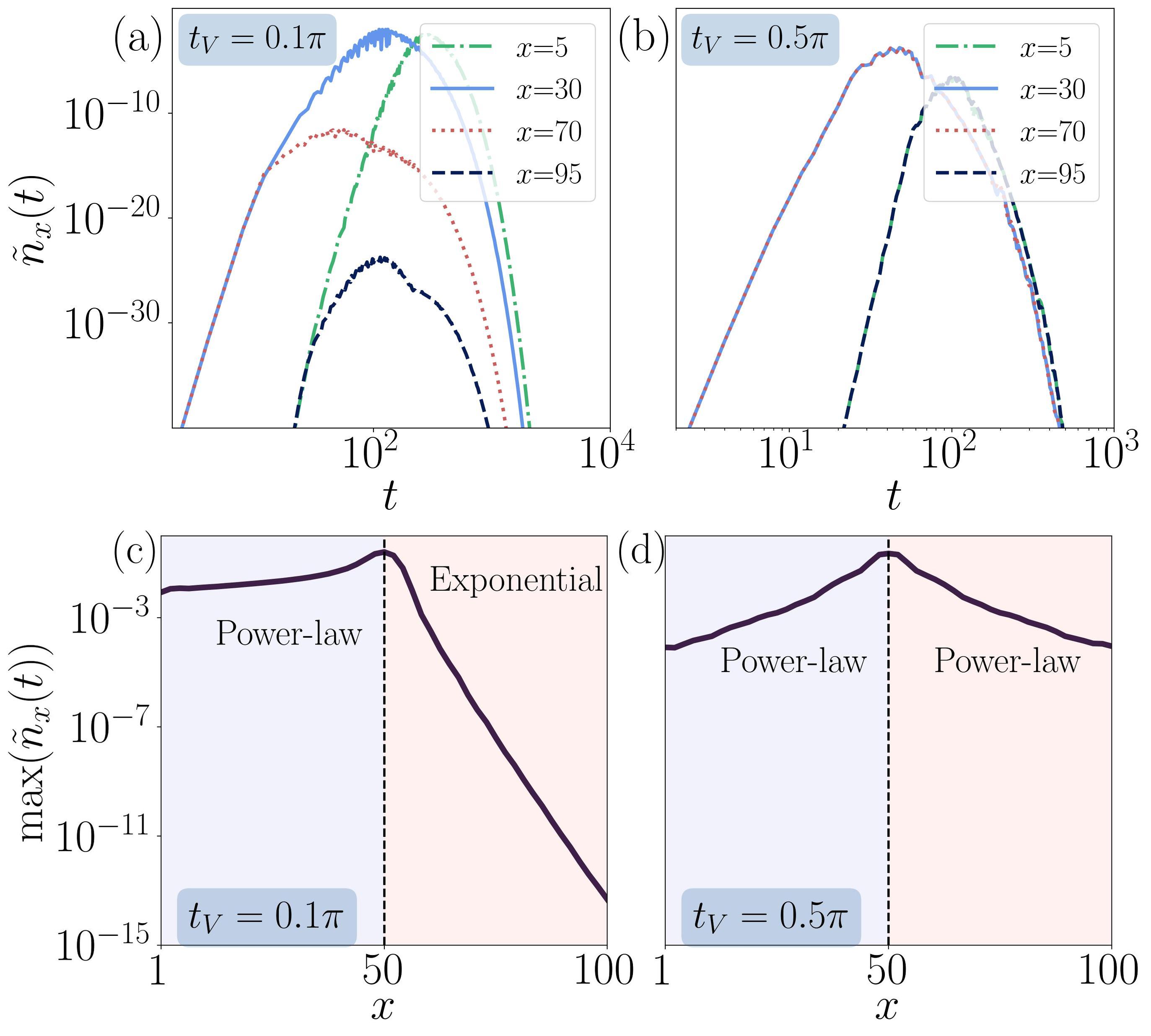}
\caption{Panels (a)–(b) show the time evolution of $\tilde{n}_x(t)$ at unit cells 
$x = 5, 30, 70,$ and $95$ for $t_V = 0.1\pi$ and $t_V = 0.5\pi$, respectively. 
Panels (c)–(d) display the maximum of $\tilde{n}_x(t)$ over the full time window 
as a function of the unit cell index. For $t_V = 0.1\pi$, the spatial profile 
reveals a pronounced information constraint: the signal decays exponentially 
toward the right, while algebraically toward the left. At the $\mathbb{Z}_2$ 
point $t_V = 0.5\pi$, this asymmetry is lifted, yielding a bipolar skin effect 
with symmetric information flow. Here, we have chosen $L=100$, $t_D = t_H = 0.5$, 
$\gamma_l = \gamma_g = 0.3$, $T = 2$.
}
\label{fig:S1}
\end{figure}
\section*{S4: Breaking the Information constraint}
A distinctive consequence of the Liouvillian skin effect is the emergence of an information constraint, arising from the unequal decay rates of the excitation amplitudes propagating in mutually opposite directions. This asymmetry effectively restricts information flow to a preferred direction. In order to further examine this behavior, we initialize the system in a zero-momentum state, corresponding to a single occupation at the central site, and monitor the spatiotemporal evolution of the relative particle number, defined as,
\begin{equation}
    \tilde{n}_x(t) = n_x (t) - n_x (\infty)
\end{equation}
Fig. \ref{fig:S1}(a) shows the evolution of $\tilde{n}_x(t)$ at representative sites, $x = 5, 30, 70$ and $95$ respectively corresponding to $t_V = 0.1\pi$ where all the wavefunctions are left localized. For each site, we extract the maximum value of $\tilde{n}_x(t)$ over the full time window. This quantity $\text{max}(\tilde{n}_x)$ turns out to be a smooth, single-valued function of $x$, whose spatial profile is shown in Fig. \ref{fig:S1}(c). The resulting damping wavefront exhibits a striking directional bias, in the sense that the signal amplitude decays exponentially toward the left $(x \rightarrow x-N/2)$ but only algebraically toward the right ($x \rightarrow x+N/2$). This asymmetric relaxation constitutes an \textit{information constraint}, where the propagation of excitations is restricted to one direction, quantitatively captured by, $I_c \propto |r|^{4x}$ \cite{information_constraint1,information_constraint2}.
\par Moreover, periodic driving provides a powerful technique to manipulate, and even reverse, this directional restriction. By tuning the drive parameter $t_V$ the unipolar skin effect undergoes a drive-induced unipolar–bipolar transition, enabling controlled inversion of the preferred direction of information flow (for example, the bias flips at $t_V =0.9\pi$). Remarkably, at the emergent $\mathbb{Z}_2$ points (in our case $t_V =0.5\pi$), the directional asymmetry collapses entirely. Here, the damping becomes helical, displaying symmetric decay in both directions, as illustrated in Fig. \ref{fig:S1}(b) and (d). This marks a complete lifting of the information constraint. Thereby, periodic driving provides a controlled means to modulate, and at special points, eliminate the directional information bias inherent to dissipative systems.
\begin{figure}[t]
         \includegraphics[width=\columnwidth]{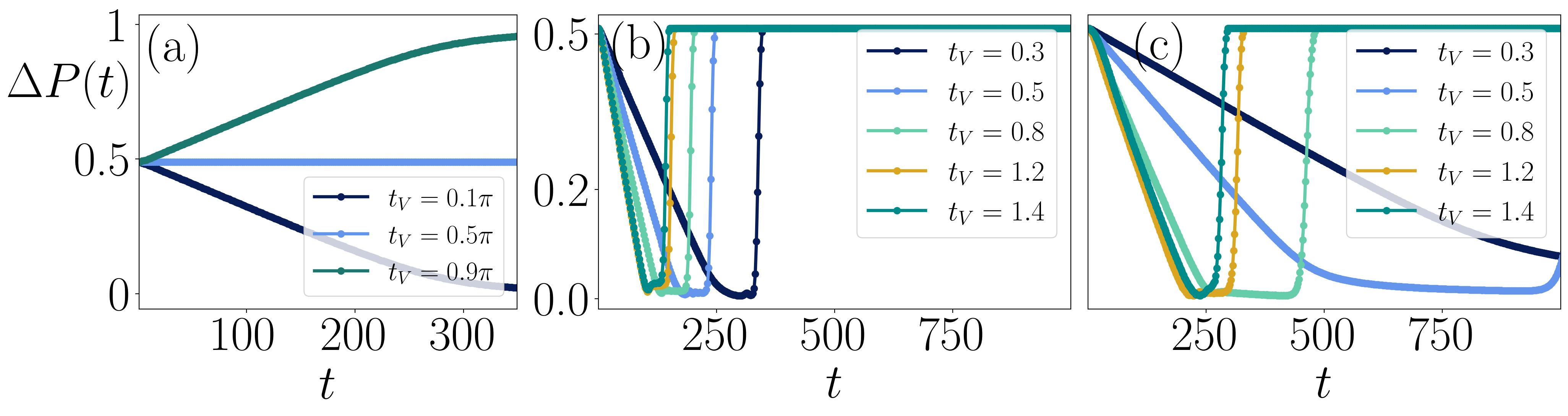}
\caption{(a) Time evolution of the dynamical polarization $\Delta P(t)$ for the 
periodically driven Liouvillian, initialized from a uniformly filled state with 
$\Delta P(0)=0.5$. For $t_V = 0.1\pi$ ($0.9\pi$), the polarization saturates near 
$0$ ($1$) before eventually relaxing to its steady-state value, while at the 
$\mathbb{Z}_2$ point $t_V = 0.5\pi$ the bi-directional skin effect keeps 
$\Delta P(t)$ pinned around $0.5$. 
(b)–(c) Time dependence of $\Delta P(t)$ for the static and driven cases, 
respectively, for several representative $t_V$ values. Although $t_V=0.3, 0.5,$ 
and $0.8$ lie in the topological regime and $t_V=1.2$ and $1.4$ in the trivial one, 
the static Liouvillian exhibits nearly identical relaxation times across all 
cases. In contrast, periodic driving leads to sharply separated relaxation 
timescales, with topological and trivial phases evolving distinctly. Here, we have chosen $L=100$, $t_D=t_H=0.5$, $\gamma_l=\gamma_g=0.3$, $T=2$.
}
\label{fig:S2}
\end{figure}
\section*{S5: Dynamical Polarization: Static vs Time-Periodic Liouvillian}
We now examine how dynamical polarization serves as a sensitive probe for distinguishing non-trivial phases in the driven Liouvillian. In the static limit, under periodic boundary conditions, the global site-averaged fermion density is defined as,
\begin{equation}
    \tilde{n}(t) = \sqrt{\sum \tilde{n}_x^2/L}
\end{equation}
discriminates the topological and trivial phases through algebraic versus exponential decay. However, this contrast collapses under open boundaries, where all the modes decay exponentially, eliminating any dynamical signature of topology. Periodic driving, however, reinstates clear topological signatures even in finite systems, motivating the need for a more robust (and unambiguous) dynamical observable. To this end, we introduce the dynamical polarization,
\begin{equation}
    \Delta P(t) = \frac{\sum_j j \, n_j(t)}{\sum_j N \, n_j(t)},
\end{equation}
which tracks the center-of-mass motion of the fermionic density and directly encodes chiral damping. Starting from a uniformly filled state with $\Delta P(t) = 1/2$ the time evolution of $\Delta P(t)$ reflects the asymmetric accumulation of particles toward one boundary. As shown in Fig. \ref{fig:S2}(a), if all Floquet eigenmodes are right-localized, $\Delta P(t)$ grows toward unity, while left-localized modes drive it toward zero; both eventually relax to the steady-state configuration at long time scale. While, in the static Liouvillian, this relaxation proceeds over nearly identical timescales for both topological and trivial phases [Fig. \ref{fig:S2}(b)], consistent with the uniform exponential decay observed in $\tilde{n}(t)$. Under periodic driving, however, the relaxation times differ sharply across phases [Fig. \ref{fig:S2}(c)], which turns out to be direct consequence governed by the drive-rescaled bulk localization length (which enters the scaling of $\Delta P(t)$ through $n(t) \propto e^{\mathcal{O}(L/\xi)}$). Moreover, as shown in Fig. \ref{fig:2}(e) of the main text, $\xi_{\text{driven}}$ remains much smaller than its static counterpart over a broad range of $t_V$, leading to stronger spatial confinement of skin modes. Consequently, $\Delta P(t)$ stays pinned near its extremal values (0 or 1) for extended durations in the topological phase, while relaxing rapidly in the trivial one. We further exploit this behavior to define a drift in the polarization (see Eq. \ref{eq:drift} of the main text), which quantitatively discriminates between the driven topological and trivial regimes, and corroborates the spectral topology obtained in Fig. \ref{fig:3} of the main text.

\end{document}